\documentclass[preprint,12pt]{elsarticle}

\usepackage{amssymb}

\usepackage{algorithm}
\usepackage{algpseudocode}

\makeatletter
\def\ps@pprintTitle{%
  \let\@oddhead\@empty
  \let\@evenhead\@empty
  \let\@oddfoot\@empty
  \let\@evenfoot\@oddfoot
}
\makeatother

 \biboptions{sort&compress,numbers}

\begin{document}

\begin{frontmatter}

\title{Development of Lightning Simulator}

\author{Shigeki Uehara}
\author{Nobuaki Shimoji}

\address{Department of Electrical and Electronics Engineering, University of the Ryukyus, 1 Senbaru, Nishihara, Okinawa, 903-0213, Japan}

\begin{abstract}
%%%
  In this work we have developed a lightning simulator.
%%%
  Reproducing the light of lightning, we adopted seven waveforms that contains two light waveforms and five current waveforms of lightning.
%%%
  Furthermore, the lightning simulator can calibrate the color of the simulated light output based on a correlated color temperature (CCT).
%%%
  The CCT of the simulated light can range over from $4269$ K to $15042$ K.
%%%
  The lightning simulator is useful for the test light source of the optical/image sensor of the lightning observation.
%%%
  Also for science education (e.g. physics education and/or earth science education etc.), the lightning simulator is available.
%%%
  In particular, the temporal change can be slowly seen by expanding the time scale from microsecond to second.
%%%
\end{abstract}


\end{frontmatter}

%\linenumbers

% Introduction
% Section 1 : Introduction
\section{Introduction}
\label{sec:introduction}

%%%
Measurement of lightning is difficult.
%%%
It is well known that lightning is very high-speed natural phenomena.
%%%
We also cannot forecast the location occuring lightning events.
%%%
For the above reason, the direct measurement of lightning is very difficult and lightning is, in general, observed indirectly.
%%%

%%%
The analysis of light emitted from a lightning channel is indirect measurements.
%%%
It is considered that the light emitted from a lightning channel contains many interior information.
%%%
This encourages to develop the optical sensor for lightning.
%%%
In development of the optical sensor, the accurate test light source is needed.
%%%
Thus in this work we have developed the test light source and we refer this as the lightning simulator.
%%%

%%%
We have developed the lightning simulator which consists of a waveform generating unit, an amplifying unit, and a light-emitting unit.
%%%
In the development, we used electronic parts realizing high-speed operation.
%%%
We adopted seven waveforms of lightning and confirmed that the lightning simulator emitted the simulated light preserving waveforms of lightning.
%%%
While the lightning simulator was developed for the test light source of the optical/image sensor, the application for science education or a lightning projector of a planetarium to simulate lightning can also be considered.
%%%

% Section 2: Instruments
\section{Instruments}
\label{sec:instrument}

\begin{figure}[h!]
  \begin{center}
    \includegraphics[width=80.0mm]{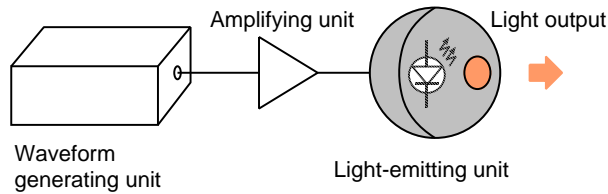}
    \caption{Schematic illustration of the lightning simulator. The lightning simulator consists of three units, the waveform generating unit, the amplifying unit, and the light-emitting unit. The integrating sphere is used in the light-emitting unit.}
    \label{fig:lightning_simulator}
  \end{center}
\end{figure}

%%%
We have developed a lightning simulator shown in Fig.~\ref{fig:lightning_simulator}.
%%%
The lightning simulator consists of the waveform generating unit, the amplifying unit, and the light-emitting unit as shown in Fig.~\ref{fig:lightning_simulator}.
%%%
The generation of the simulated light of lightning is as follows:
%%%
creating a lightning waveform in the waveform generating unit,
%%%
regulating the lightning waveform to appropriate value in the amplifying unit,
%%%
and emitting the light preserving the simulated waveform in the light-emitting unit.
%%%
In the light-emitting unit, although several LEDs are used, the light is uniformized in the integrating sphere.
%%%
% Figure

%%%
The details of the units above are shown in Fig.~\ref{fig:circuits}.
%%%
% Figure
\begin{figure*}[h!]
  \begin{center}
    \includegraphics[width=90.0mm]{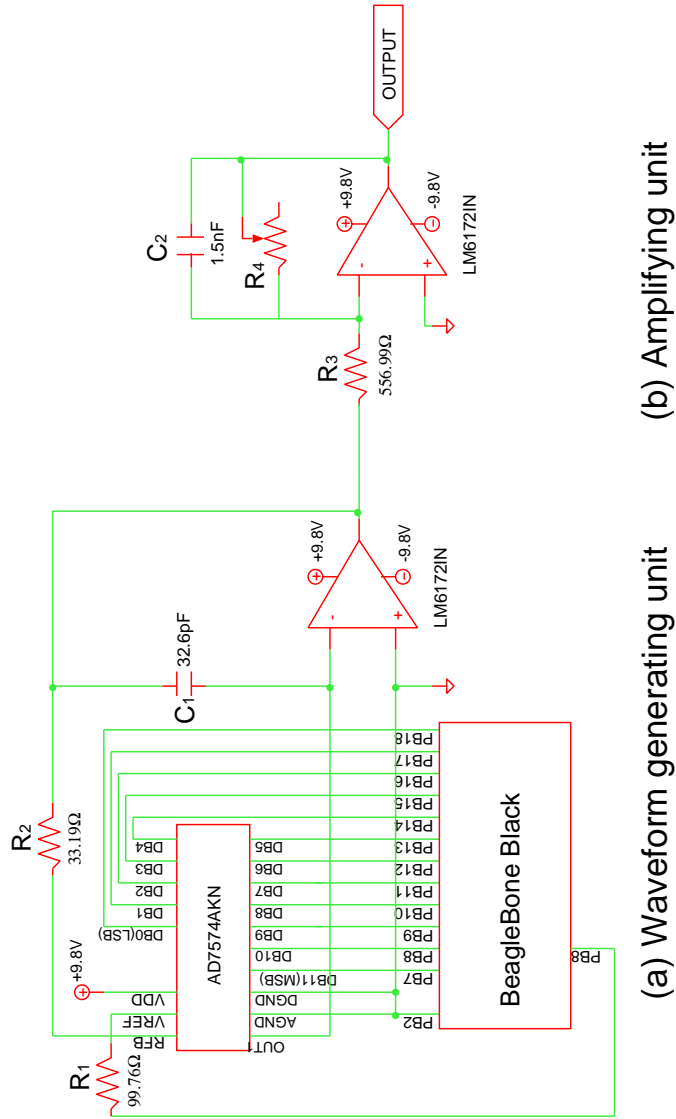}
    \caption{Circuits of the (a) waveform generating, (b) amplifying, and (c) light-emitting unit. The external voltage reference VDD of the DAC is $9.8$ V. The reference voltage VREF of the DAC is $5$ V. The supply voltage for the Op-Amp is $\pm 9.8$ V. The output waveforms of the amplifying unit is transfered to the input, INPUT, of the light-emitting unit.}
    \label{fig:circuits}
  \end{center}
\end{figure*}
%%%
\addtocounter{figure}{-1}
%%%
% Figure
\begin{figure*}[h!]
  \begin{center}
    \includegraphics[width=100.0mm]{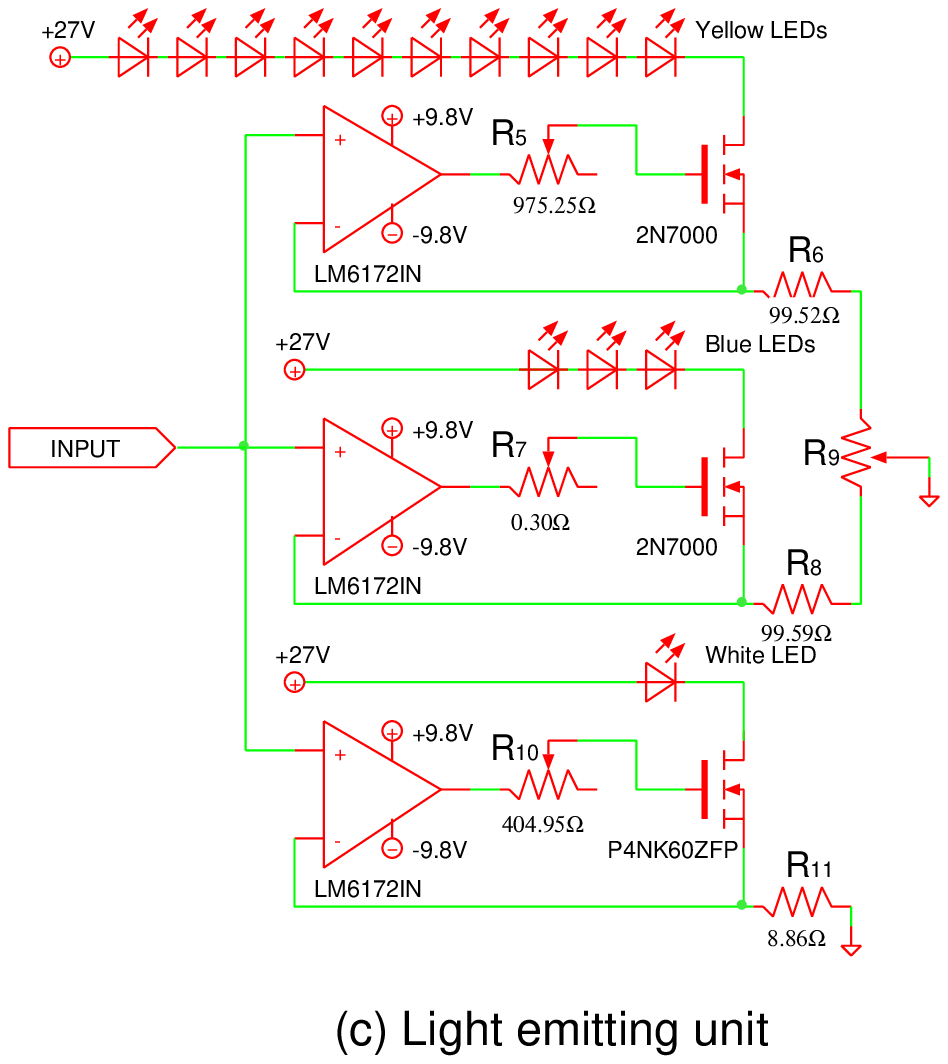}
    \caption{(Continued)}
    \label{fig:circuits}
  \end{center}
\end{figure*}
%%%
Reproducing high-speed lightning waveforms, we used electronic parts (e.g. operational amplifier (Op-Amp), n-channel enhancement MOS FETs, and LEDs) having nanosecond respons speed.
%%%
In the waveform generating unit (Fig.~\ref{fig:circuits} (a)), binary signals of the simulated waveform of lightning are created by a microcontroller (BeagleBorn Black, Rev A5A) in which Debian/Gnu Linux $7$ (kernel $3$.$8$.$13$ - bone $28$) for the operating system and gcc ver.$4$.$6$.$3$ (debian $4$.$6$.$3$-$14$) for the C compiler are adopted.
%%%
Since, the $10$/$90$ rise time is inconstant for each lightning, we set appropriate value in $1.0 \mu$s $\le \Delta t \le 2.4$ $\mu$s to $\Delta t$ at the first step.
%%%
At the other steps, $\Delta t = 2.4 \mu$s.
%%%
We show the pseudocode of the waveform generation algorithm to Algorithm~\ref{algorithm:generating_waveforms}.
% Algorithm (Reproducing waveforms)
\begin{algorithm}
\caption{Pseudocode of the waveform generation algorithm.}
\label{algorithm:generating_waveforms}
\begin{algorithmic}
%%%
\State // (Variable Declaration)
%\State
\State // $GPIO pin$: I/O pin on the microcontroller (BeagleBone Black)
\State // $Time Step$: Time step $\Delta t$ ($\mu$s).
\State // $Waveform$[]: array containing the waveform data.
\State // $t_{max}$: Maximum number of the array of the waveform data.
\State
\State Initializing $GPIO pin$
\State
\For{$Time \le t_{max}$ }
\State $Time \leftarrow Time + Time Step$
\State $GPIO pin \leftarrow Waveform(Time)$
\State Output($GPIO pin$)
\EndFor
\end{algorithmic}
\end{algorithm}
%%%
Then by the $12$ bit resolution digital-to-analog converter (DAC), AD7574AKN, the binary signals are converted to an analog waveform.
%%% 
Fig.~\ref{fig:circuits} (b) shows the amplifying unit.
%%%
The output (simulated waveform) created by the waveform generating unit (Fig.~\ref{fig:circuits} (a)) is regulated to appropriate value by the amplifying unit.
%%% 
The variable resistor $R_{4}$ is to regulate the amplification degree.
%%%
The capacitor $C_{2}$ is the compensation capacitor.
%%%
The light-emitting unit (Fig.~\ref{fig:circuits} (c)) contains a high-power white LED (typ. $103250$ mcd), three high-brightness blue LEDs (typ. $12000$ mcd), and ten high-brightness yellow LEDs (typ. $4000$ mcd).
%%%
The brightness of the blue and yellow LEDs are adaptable by regulating the variable resistor $R_{9}$ ($10$ k$\Omega $).
%%%
Regulating the brightness of the blue and yellow LEDs we can calibrate the CCT of the light output.
%%%
The brightness of the white LED is not adaptable, namely, constant.
%%%
The variable resistors $R_{5}$, $R_{7}$, and $R_{10}$ in Fig.~\ref{fig:circuits} (c) are to prevent oscillation.
%%%
In the light-emitting unit, uniforming the light emitted from the white, blue, and yellow LEDs the integrating sphere is used.
%%%
The diffuse reflectance of the diffuse reflectance material coated in the integrating sphere are shown in Fig.~\ref{fig:reflectance}\cite{TMITRI}.
%%%
% Figure
\begin{figure}[h!]
  \begin{center}
    \includegraphics[width=80.0mm]{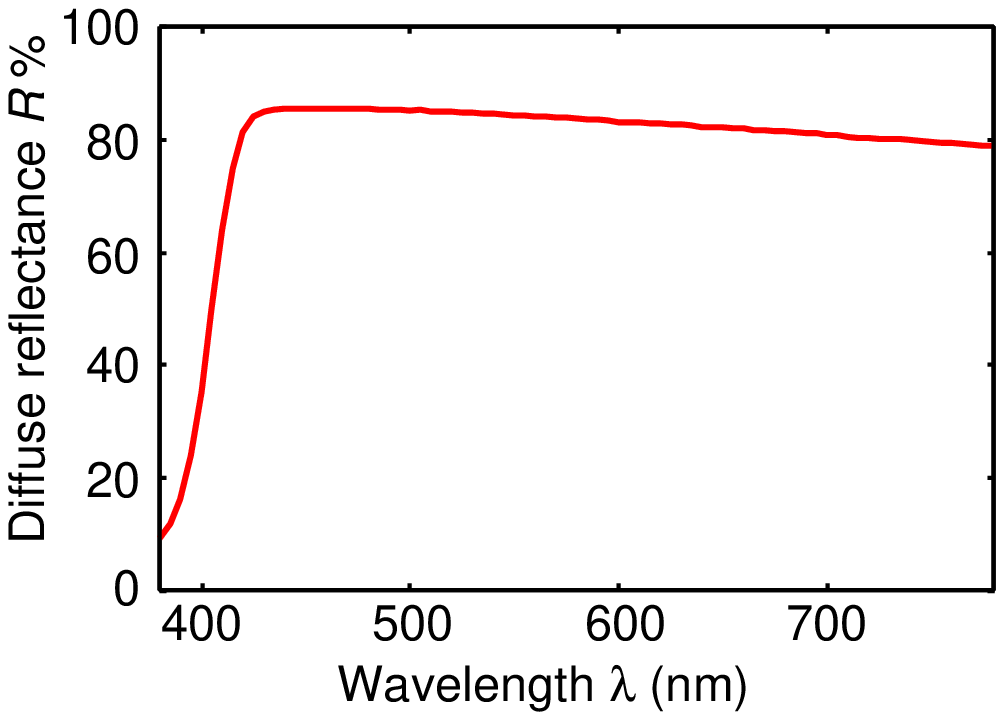}
    \caption{Diffuse reflectance of the diffuse reflectance materials of the integrating sphere. The data was provided by Tokyo Metropolitan Industrial Technology Research Institute~\cite{TMITRI}.}
    \label{fig:reflectance}
  \end{center}
\end{figure}
%%%
From Fig.~\ref{fig:reflectance}, it is seen that in the visible light range, the diffuse reflectance is $80$ -- $86$ \% without deep blue region.
Fig.~\ref{fig:integrating_sphere} shows t\
he side view of the integrating sphere.
%%%
% Figure
\begin{figure}[h!]
  \begin{center}
    \includegraphics[width=70.0mm]{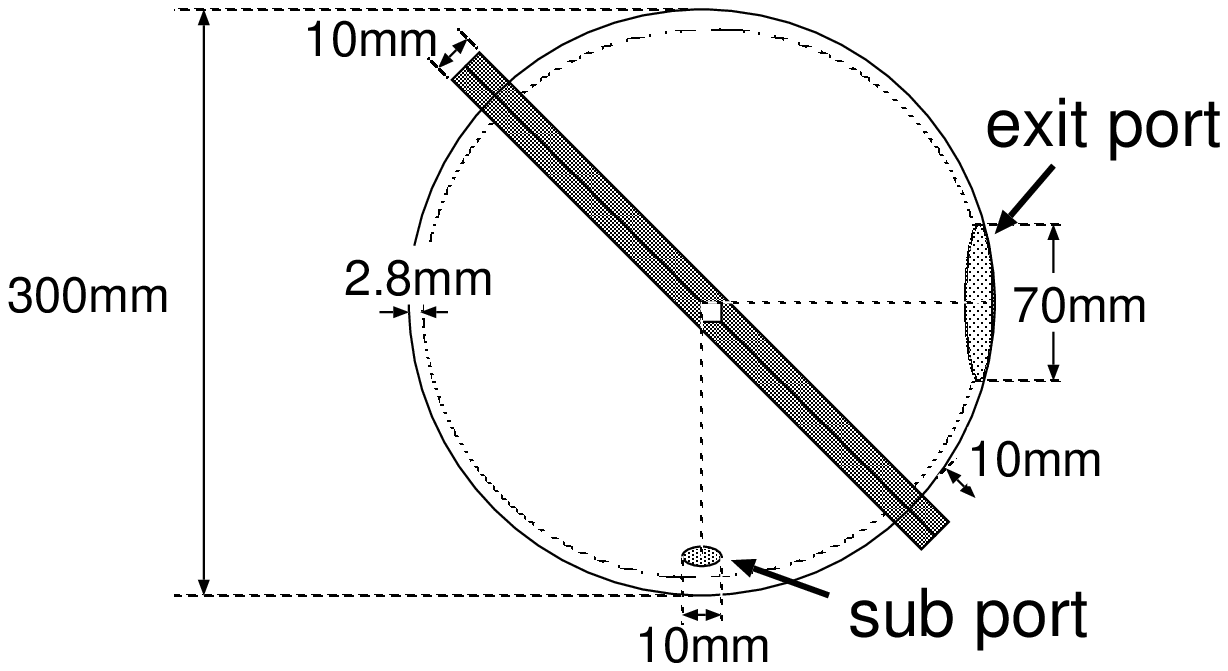}
    \caption{Side view of the integrating sphere. The outer diameter of the integrating sphere is $300$ mm and the diameter of the exit port is $70$ mm. The thickness is $2.8$ mm.}
    \label{fig:integrating_sphere}
  \end{center}
\end{figure}
%%%
The diameters of the integrating sphere and the exit port are $300$ and $70$ mm, respectively.
%%%  

% Methods
% Section 2: Methods
\section{Methods}
\label{sec:methods}

% Subsection : Performance test of the integrating sphere
\subsection{Performance test of the integrating sphere}
\label{lbl:performance_test}

\begin{figure*}[h!]
  \begin{center}
    \includegraphics[width=110.0mm]{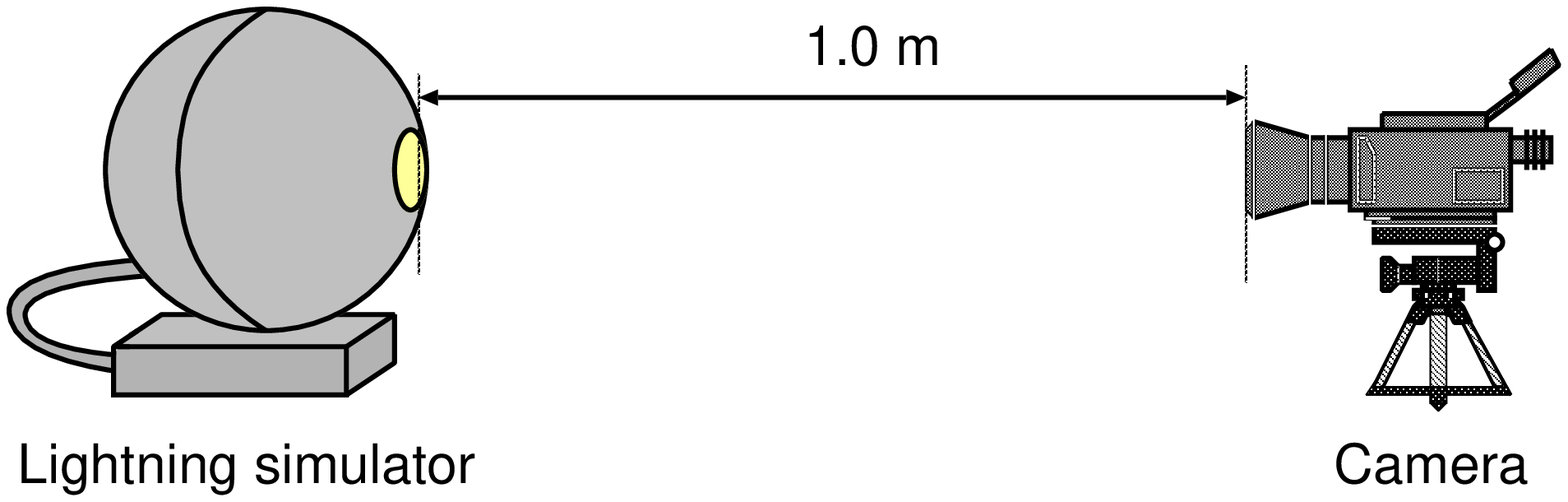}
    \caption{Experimental setup to test the uniformity of the integrating sphere and the CCT of the simulated light. The distance between the lightning simulator and the digital still camera is $1.0$ m.}
    \label{fig:condition}
  \end{center}
\end{figure*}

%%%
We have evaluated the performance of the integrating sphere.
%%%  
Fig.~\ref{fig:condition} shows the schematic illustration for the experimental setup to evaluate the performance of the lightning simulator.
%%%
%%%
The distance between the lightning simulator and the digital still camera is $1.0$ m.
%%%
The photographing condition settings are as follows: file type is ``NEF (Nikon NEF raw image)'', compression is ``Uncompressed'', F number is ``$5.6$'', ISO is ``$800$'', color space is ``sRGB''.
%%%
Under the photographing condition above, we measured the uniformity of the integrating sphere and the CCT of the simulated light.
%%% 

% Subsubsection : Uniformity of the light emitted from the integrating sphere
\subsubsection{Uniformity of the light emitted from the integrating sphere}
\label{lbl:uniformity}

\begin{figure}[h!]
  \begin{center}
    \includegraphics[width=80.0mm]{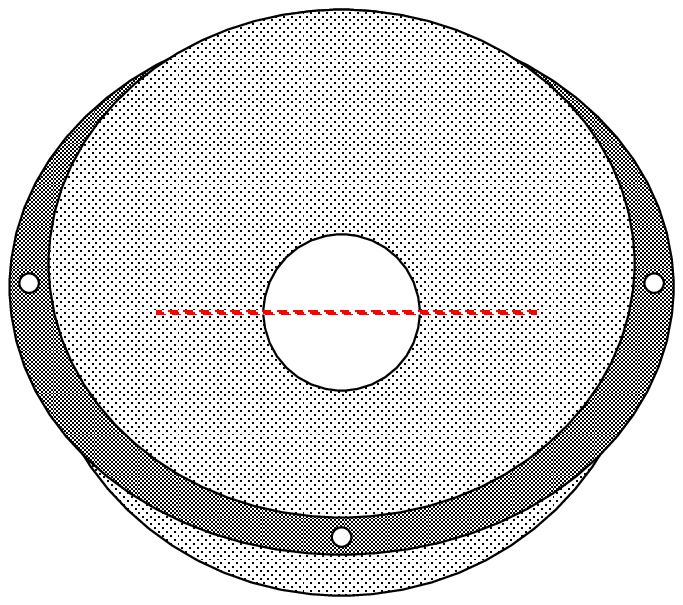}
    \caption{Front view of the integrating sphere. The dashed line acrossing the exit port is the test line to check the uniformity of light.}
    \label{fig:exit_port}
  \end{center}
\end{figure}

%%%
We analyzed the uniformity of the integrating sphere.
%%%
Fig.~\ref{fig:exit_port} shows the exit port of the integrating sphere.
%%%
%%%
The dashed line in Fig.~\ref{fig:exit_port} is to check the uniformity of light emitted from the integrating sphere.
%%%
For the measurement of the uniformity, only the high-power white LED driven by the constant current was used and the blue and yellow LEDs did not be used.
%%%
The driving current of the high-power white LED was set to $338.6$ mA.
%%%
Under these condition, we analyzed the variation of the grayscale pixel value ($256$ levels) on the test line shown by the dashed line in Fig.~\ref{fig:exit_port}.
%%% 

% Subsubsection : CCT of the lightning simulator
\subsubsection{CCT of the lightning simulator}
\label{lbl:CCT_of_the_lightning_simulator}

\begin{figure}[h!]
  \begin{center}
    \includegraphics[width=80.0mm]{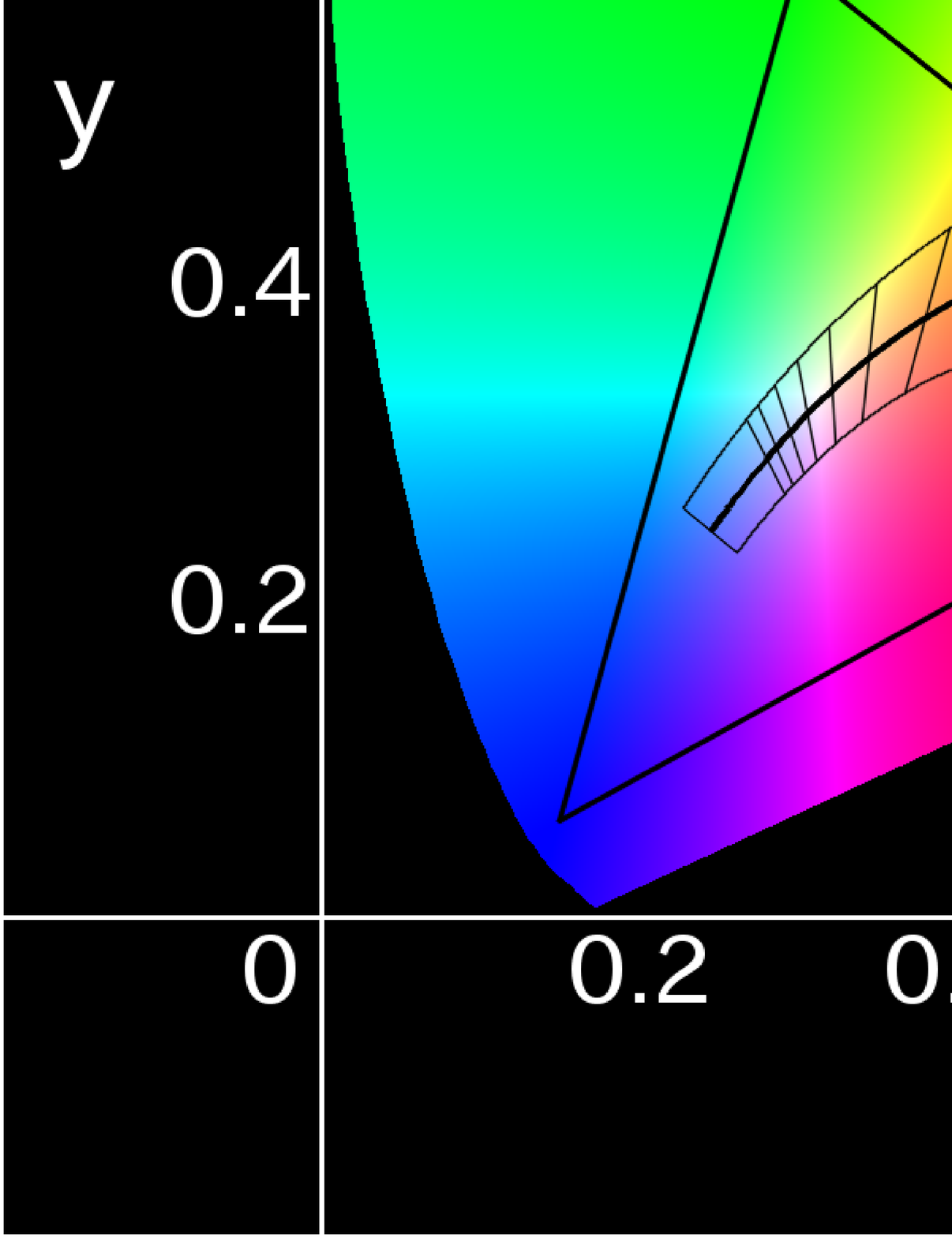}
    \caption{CIE $1931$ $xy$-chromaticity diagram\cite{sRGB}. The sRGB color triangle and the black-body locus are drawn. The thin lines crossing the black-body locus are the isotemperature lines. The thin lines along the black-body locus are the $\Delta$uv lines of $\pm 0.02\Delta$uv. }
    \label{fig:xy_chromaticity_diagram}
  \end{center}
\end{figure}

%%%
We analyzed the CCT of the lightning simulator.
%%%
All LEDs in the light-emitting unit were driven by the constant current.
%%%  
As mentioned in the previous section, the luminous intensity of the blue and yellow LEDs are adaptable by regulating the variable resistor $R_{9}$.
%%%
Thus, the CCT of the lightning simulator was calibrated regulating the variable resistor $R_{9}$.
%%%
For the evaluation of the CCT, all LEDs were driven by the five test conditions TC$1$ -- TC$5$ shown in Table~\ref{tbl:current_conditions}.
%%%
\begin{table*}[ht]
  \begin{center}
  \caption{Five test conditions. The current values are defined for each test conditions TC$1$ -- TC$5$.}
  \label{tbl:current_conditions}
  \begin{tabular}{cccc}
%  \begin{table}{c|c}
    \hline
    Condition ID & \multicolumn{3}{l}{LED driving current (mA)} \\
    \cline{2-4}
    & white & blue & yellow \\
    \hline\hline
    TC$1$ & $338.6$ & $0.296$ & $29.800$\\
    TC$2$ & $338.6$ & $0.299$ & $15.000$\\
    TC$3$ & $338.6$ & $0.588$ & $0.585$\\
    TC$4$ & $338.6$ & $15.000$ & $0.299$\\
    TC$5$ & $338.6$ & $29.800$ & $0.296$\\
    \hline
  \end{tabular}
  \end{center}
\end{table*}
%%% 
Usually, the CCT is discussed on the CIE $1931$ $xy$-chromaticity diagram shown in Fig.~\ref{fig:xy_chromaticity_diagram}.
%%%
%%%
Thus we analyze the CCT of the light emitted from the lightning simulator by plotting to the $xy$-chromaticity diagram.
%%%

% Subsection : Test of Lightning Simulator
\subsection{Test of the simulated light output of the lightning simulator}
\label{lbl:test_of_lightning_simulator}

%%%
Reproducing the simulated light of the lightning, we adopted seven lightning waveforms shown in Table~\ref{tbl:lightning_waveforms}.
%%%
Since it is suggested that there are the strong positive correlation between the light intensity and the channel current of lightning\cite{Idone_Orville,Gomes_Cooray,Wang,Zhou_et_al}, in this work we proactively used the waveforms of the channel current as the waveforms of the simulated light of the lightning simulator.
%%%
\begin{table*}[ht]
%\begin{tabular}
  \begin{center}
  \caption{Waveforms used in the lightning simulator. The ID, WF$1$ and WF$7$, are the light signal waveforms and WF$2$ -- WF$6$ are the lightning current waveforms. The $10
$/$90$ rise time and the duration of wave tail of WF$7$ were indeterminable since the waveform of WF$7$ is very complex. The ``Type'' indicates the waveform type either a light signal or a lightning current.}
  \label{tbl:lightning_waveforms}
  \begin{tabular}{llllll}
    \hline
    ID & \multicolumn{2}{l}{Duration ($\mu$s)} & Type & Fugire number & Reference \\
    \cline{2-3}
    & $10$/$90$ rise time & wave tail & Light/Current & \\
    \hline\hline
    WF$1$ & $0.9$ & $4.0$ & Light & Fig.~$2$ & Ref.~\cite{Wang}\\
    WF$2$ & $1.8$ & $18.0$ & Current & Fig.~$2$ (RS $1$) & Ref.~\cite{Zhang}\\
    WF$3$ & $8.1$ & $109.1$ & Current & Fig.~$12$ & Ref.~\cite{Berger}\\
    WF$4$ & $1.07$ & $45.2$ & Current & Fig.~$13$ & Ref.~\cite{Berger}\\
    WF$5$ & $7.25$ & $134.9$ & Current & Fig.~$2$ a & Ref.~\cite{SilvrioVisacro}\\
    WF$6$ & $18.75$ & $94.8$ & Current & Fig.~$1$ a & Ref.~\cite{Miguel}\\
    WF$7$ & $-$ & $-$ & Light & Fig.~$6$ & Ref.~\cite{Miguel}\\
    \hline
  \end{tabular}
  \end{center}
\end{table*}
%%%
Table~\ref{tbl:lightning_waveforms} shows the waveform information used in the lightning simulator.
%%%
The WF$1$~\cite{Wang} and WF$7$~\cite{Miguel} in Table~\ref{tbl:lightning_waveforms} are the waveforms of the light signal of lightning and the WF$2$ -- WF$6$\cite{Zhang,Berger,SilvrioVisacro,Miguel} in Table~\ref{tbl:lightning_waveforms} are the current waveforms of lightning.
%%%

\begin{figure*}[h!]
  \begin{center}
    \includegraphics[width=80.0mm]{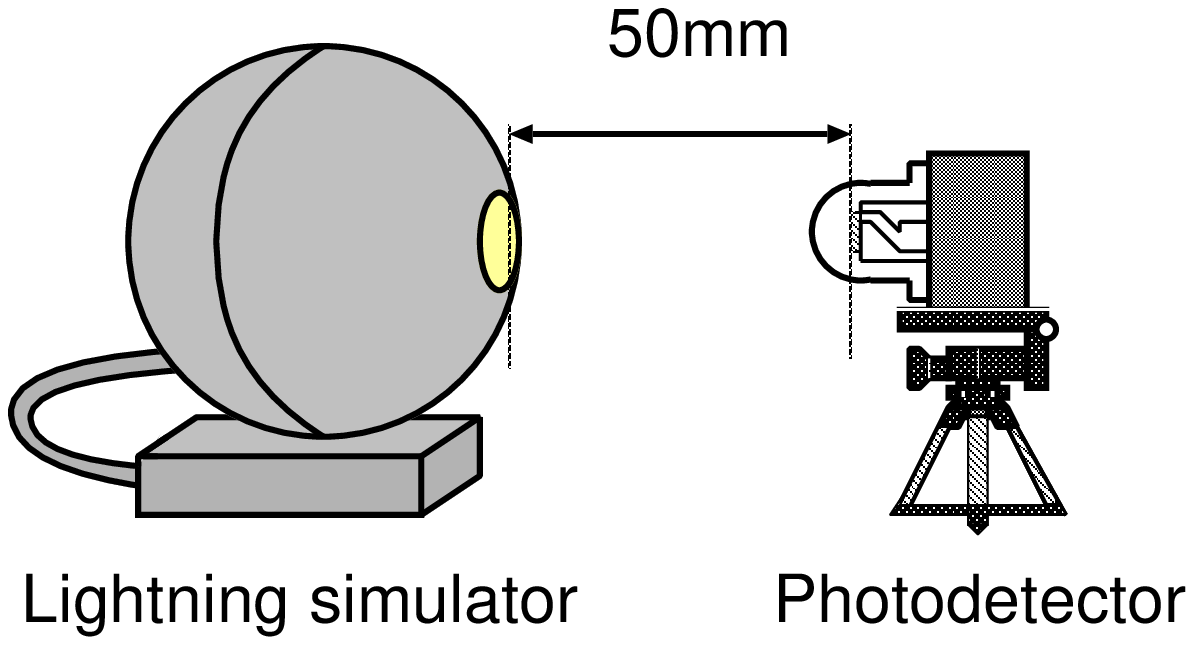}
    \caption{Schematic illustration for measurement of the light output. The distance between the lightning simulator and the photodetector are $50$ mm.}
    \label{fig:measurement_by_photodiode}
  \end{center}
\end{figure*}

%%%
Using these waveforms, we carried out the test of the simulated light output of the lightning simulator.
%%%
The light emitted from the lightning simulator was measured by using the photodetector as shown in Fig.~\ref{fig:measurement_by_photodiode}.
%%%
%%% 
The distance between the photodetector and the exit port is $50$ mm.
%%%
Under the photographing setup above, the test of the lightning simulator was carried out.

% Results
% Section 3: Results
\section{Results}
\label{sec:results}

% Subsection: Performance test of the integrating sphere 
\subsection{Performance test of the integrating sphere}
\label{sec:results_performance_test}

\begin{figure}
 \begin{center}
   \includegraphics[width=80.0mm]{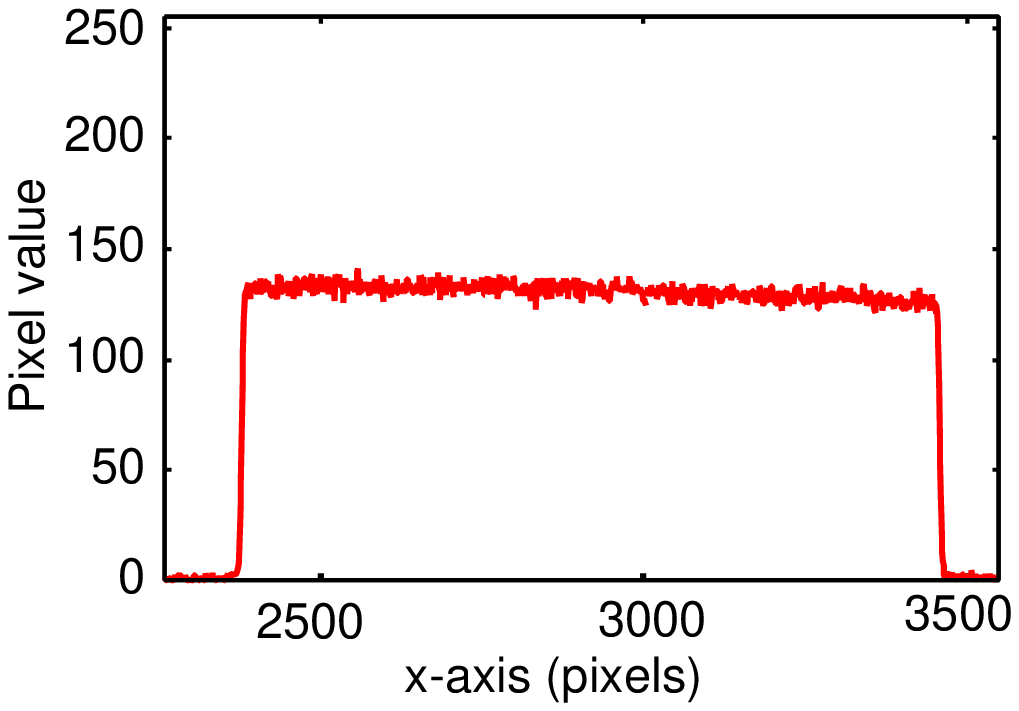}
   \caption{Variation of the grayscale pixel value ($256$ levels) on the dashed line drawn on Fig.\ref{fig:exit_port}.}
   \label{fig:results_uniformity}
 \end{center}
\end{figure}

%===< Uniformity >===% 
%%%
Fig.~\ref{fig:results_uniformity} shows the variation of the grayscale pixel value ($256$ levels) on the dashed line in Fig.~\ref{fig:exit_port}.
%%%
We can see that the pixel values on the exit port is holding the constant value approximately.
%%%
This means that the light emitted from the lightning simulator is uniform.
%%%

\begin{figure}
 \begin{center}
   \includegraphics[width=40.0mm]{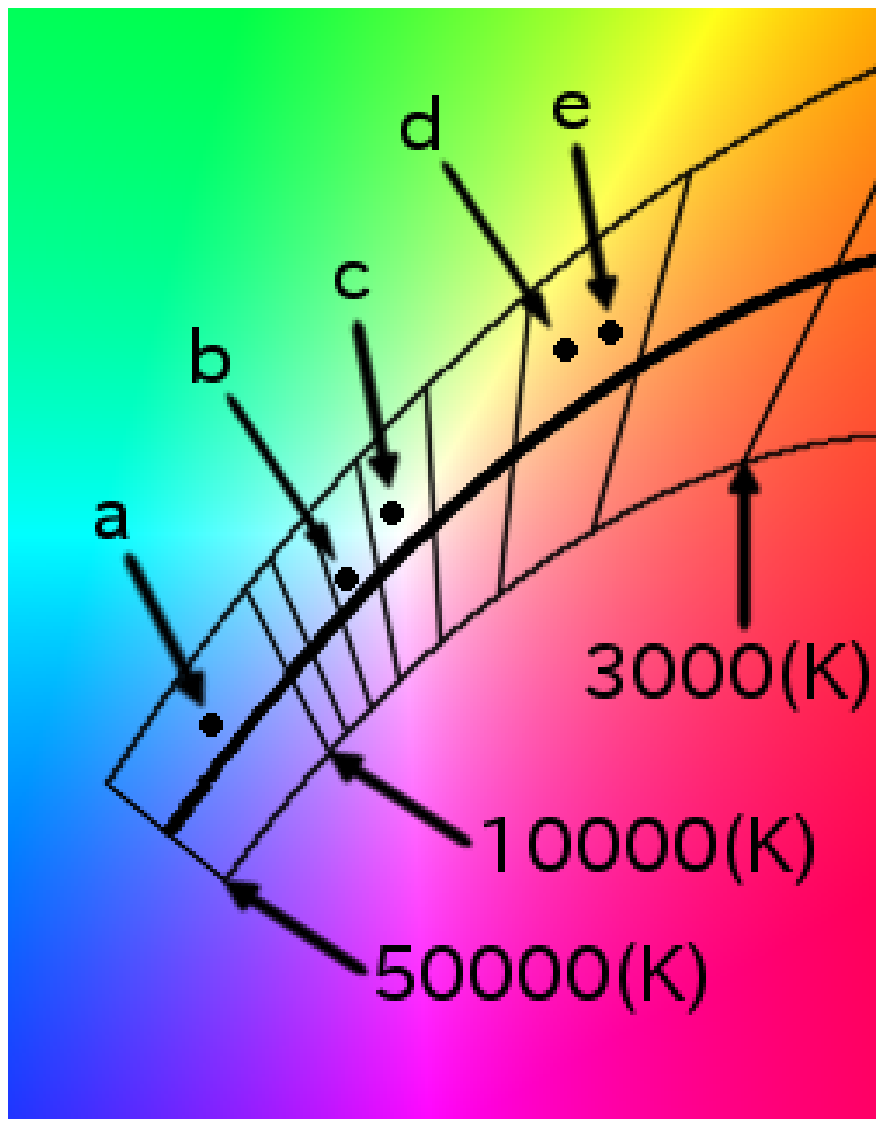}
   \caption{CCT of the simulated light emitted from the lightning simulator. The points a -- e indicate the CCT obtained under the conditions TC$1$ -- TC$5$, respectively. The thick line is the black-body locus. The thin lines which is acrossing with the black-body locus is the isotemperature lines. The thin lines along the black-body locus are the $\Delta$uv lines of $\pm 0.02\Delta$uv.}
   \label{fig:results_cct}
 \end{center}
\end{figure}

%%%
Fig.~\ref{fig:results_cct} shows the magnified figure of the $xy$-chromaticity diagram on which the simulated lights for the five test conditions (see Table~\ref{tbl:current_conditions}) are plotted.
%%%
The points a -- e on Fig.~\ref{fig:results_cct} were obtained under the test condition TC$1$ -- TC$5$.
%%%
We confirmed that the CCT of the simulated light ranges over from $4269$ K to $15042$ K with regulating the variable resistor $R_{9}$.
%%%
From Fig.~\ref{fig:results_cct}, it is seen that the CCT a -- e of the lightning simulator are $15042$, $7578$, $6597$, $4618$ and $4269$ K under the test condition TC$1$ -- TC$5$.
%%% 
From above, we can find that the CCT of the simulated light changes on the line segment a -- e as regulating the variable resistor $R_{9}$.

% Subsection: Test of the lightning simulator                                      
\subsection{Test of the simulated light output of the lightning simulator}
\label{sec:results_test_of_lightning_simulator}

\begin{figure*}
 \begin{center}
   \includegraphics[width=130.0mm]{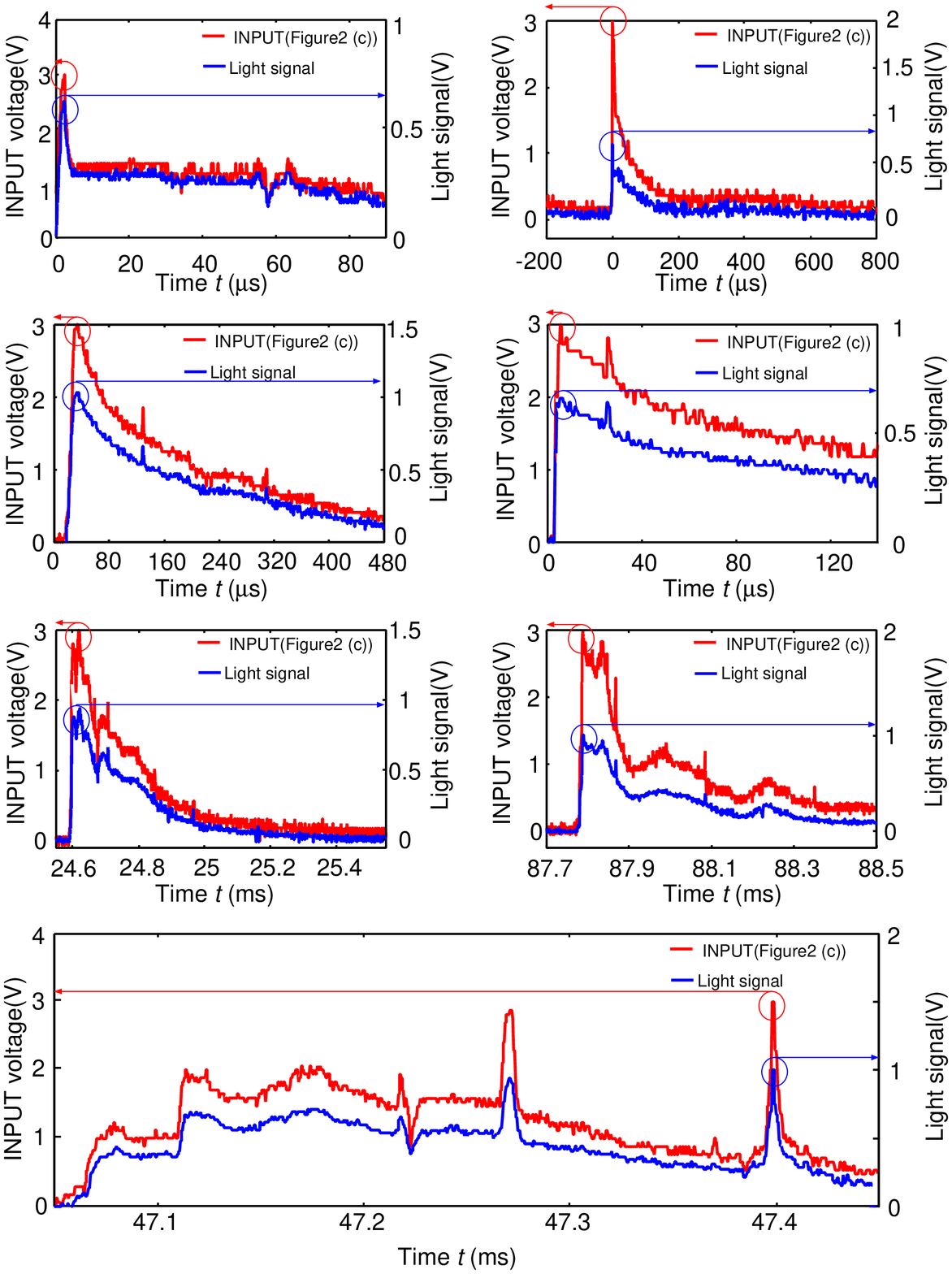}
   \caption{Waveforms of both the input voltage of the light-emitting unit and the simulated light emitted from the lightning simulator. The INPUT voltage indicate the input of the light-emitting unit. The light signal indicate the waveform detected by the photodetector.}
   \label{fig:results_waveforms}
 \end{center}
\end{figure*}

%%%
Fig.~\ref{fig:results_waveforms} shows the input waveforms of the light-emitting unit and the simulated light waveforms detected by the photodetector.
%%%
The $10$/$90$ rise time and the duration of wave tail are summarized in Table~\ref{tbl:results_wave_front_and_wave_tail}.
%%%
Fig.~\ref{fig:results_waveforms} indicate that the simulated light can reproduce the drastic change.
%%%
\begin{table*}[ht]
  \begin{center}
  \caption{Duration of the $10$/$90$ rise time and wave tail of the results in Fig.~\ref{fig:results_waveforms}.}
  \label{tbl:results_wave_front_and_wave_tail}
  \begin{tabular}{lllll}
    \hline
    Figure & \multicolumn{2}{l}{$10$/$90$ rise time ($\mu$s)} & \multicolumn{2}{l}{wave tail ($\mu$s)} \\
    \cline{2-5}
    & Input voltage & Light signal & Input voltage & Light signal \\
    \hline\hline
    Fig.~\ref{fig:results_waveforms} (a) & $1.1$ & $1.2$ & $3.7$ & $3.9$ \\
    Fig.~\ref{fig:results_waveforms} (b) & $1.0$ & $2.0$ & $20.0$ & $21.0$ \\
    Fig.~\ref{fig:results_waveforms} (c) & $10.4$ & $10.2$ & $100.8$ & $115.2$ \\
    Fig.~\ref{fig:results_waveforms} (d) & $1.6$ & $1.2$ & $92.4$ & $108.4$ \\
    Fig.~\ref{fig:results_waveforms} (e) & $6.4$ & $6.2$ & $139.6$ & $154.0$ \\
    Fig.~\ref{fig:results_waveforms} (f) & $6.8$ & $6.2$ & $82.25$ & $96.2$ \\
    Fig.~\ref{fig:results_waveforms} (g) & $-$ & $-$ & $-$ & $-$ \\
    \hline
  \end{tabular}
  \end{center}
\end{table*}
%%%

%%%
%%%

%%%
%%%

%%%

% Discussion
% Section 4: Discussion
\section{Discussion}
\label{sec:discussion}

%%%
We have developed the lightning simulator that emit the uniform light.
%%%
The color of the light emitted from the lightning simulator can calibrate based on the CCT.
%%% 
It is already reported that the lightning opacity is thin\cite{Uman_and_Orville}, namely not the blackbody radiation.
%%%
However, we obtained results(Aoyama and Shimoji, 2013, unpublished data) that when we plot the image pixels of two lightning images to the CIE$1931$ $xy$-chromaticity diagram, the points distribute mainly in the $\pm 0.02\Delta$uv lines.
%%%
This suggests that the color of lightning can be define by the CCT.
%%%
Thus, we designed the lightning simulator which is adaptable the color based on the CCT, since the CCT can define the color quantitatively.
%%%

%%%
It was also confirmed that the lightning simulator reproduce the original lightning waveforms.
%%% 
Idone and Orville\cite{Idone_Orville}, Gomes and Cooray\cite{Gomes_Cooray}, Wang {\it et al.}\cite{Wang}, and Zhou {\it et al.}\cite{Zhou_et_al} suggest that there exists a strong positive correlation between the current and light signal of lightning channel.
%%%
Therefore we used the current waveforms of lightning as the light waveform.
%%% 

%%%
The lightning simulator is useful for the test light source to develop the optical/image sensor.
%%%
Hereafter, it will be considered that a ground-based observation of lightning and/or space-based observation of planetary lightning increase.
%%%
Accordingly, high-speed optical/image sensors will be developed.
%%%
A accurate test light source reproducing the lightning waveform is necessary for evaluating the optical/image sensor.
%%%
We consider that the lightning simulator developed in this work will be used as the test light source.
%%%  
It is also considered that the lightning simulator can be used for science education.
%%%
Since the microcontroller is programmable, the time scale of the simulated waveform can be expanded from the microsecond order to the second order.
%%%
Therefore we can visually observe the slow change of the simulated light.
%%%
Thus the lightning simulator is useful for science education (especially physics education and earth science education).
%%% 

% Conclusions
% Section 4: Conclusion
\section{Conclusion}
\label{sec:conclusion}

%%%
We have developed the lightning simulator.
%%%
The CCT of the light emitted from the lightning simulator can be calibrated from $4269$ K to $15042$ K.
%%%
The lightning simulator in this work can simulate seven waveforms.
%%%
It is considered that the lightning simulator is useful for the test light source for the image/optical sensor of the lightning observation.
%%%
Furthermore, we consider that expanding the time scale from microsecond order to second order, the lightning simulator can be used for science education.
%%%

% Acknowledgements
% Section 5: Acknowledgements             
\section{Acknowledgements}
\label{lbl:acknowledgements}

%%%
The authors would like to thank Yu Iida, Ryoma Aoyama, Wataru Hasegawa, and Zensei Iha for technical guidance which improved the quality of the paper.
%%%
We are thankful to Yasuhiko Teruya for the acrylic processing of the integrating sphere.
%%%
The authors would also need to acknowledge Optical Radiation and Acoustics Technology Group, Tokyo Metropolitan Industrial Technology Research Institute that provided the reflectance data of the white pigment in the integrating sphere.
%%%

\bibliographystyle{elsarticle-num}
\bibliography{bib_ls}

\begin{thebibliography}{10}
\expandafter\ifx\csname url\endcsname\relax
  \def\url#1{\texttt{#1}}\fi
\expandafter\ifx\csname urlprefix\endcsname\relax\def\urlprefix{URL }\fi
\expandafter\ifx\csname href\endcsname\relax
  \def\href#1#2{#2} \def\path#1{#1}\fi

\bibitem{TMITRI}
{Tokyo Metropolitan Industrial Technology Research Institute}, {}, {The diffuse
  reflectance data were provided by the Optical Radiation and Acoustics
  Technology Group, Tokyo Metropolitan Industrial Technology Research
  Institute.}

\bibitem{sRGB}
{IEC 61966-2-1}, {Multimedia systems and equipment - Colour measurements and
  management - Part 2-1: Colour management - Default RGB color space - sRGB
  (International Electrotechnical Commission, Geneva, 1999-10}.

\bibitem{Idone_Orville}
{V. P. Idone, R. E. Orville}, {Correlated peak relative light intensity and
  peak current in triggered lightning subsequent return strokes}, J. Geophys.
  Res. 90 (1985) 6159--6164.

\bibitem{Gomes_Cooray}
{C. Gomes and V. Cooray}, {Correlation between the optical signatures and
  current waveforms of long sparks: applications in lightning research}, J.
  Electrost 43 (1998) 267--274.

\bibitem{Wang}
{D. Wang, N. Takagi, T. Watanabe, V. A. Rakov, M. A. Uman, K. J. Rambo, M. V.
  Stapleton}, {A comparison of channel-base currents and optical signals for
  rocket-triggered lightning strokes}, Atmos. Res. 76 (2005) 412 -- 422.

\bibitem{Zhou_et_al}
{E. Zhou, W. Lu, Y. Zhang, B. Zhu, D. Zheng, Y. Zhang}, {Correlation analysis
  between the channel current and luminosity of initial continuous and
  continuing current processes in an artificially triggered lightning flash},
  Atmos. Res. 129--130 (2013) 79--89.

\bibitem{Zhang}
{Yijun Zhang, Shaojie Yang, Weitao Lu, Dong Zheng, Wansheng Dong, Bin Li,
  Shaodong Chen, Yang Zhang, Luwen Chen}, {Experiments of artificially
  triggered lightning and its application in Conghua, Guangdong, China}, Atmos.
  Res. 135–-136 (2014) 330 -- 343.

\bibitem{Berger}
{K. Berger, R. B. Anderson and H. Kr\"oninger}, {Parameters of lightning
  flashes}, {Electra} No.41 (1975) 23 -- 37.

\bibitem{SilvrioVisacro}
{Silv\'erio Visacro, Claudia R. Mesquita, Alberto De Conti, Fernando H.
  Silveira}, {Updated statistics of lightning currents measured at Morro do
  Cachimbo Station}, Atmos. Res. 117 (2012) 55 -- 63.

\bibitem{Miguel}
{Miguel Guimar\~aes, Listz Araujo, Clever Pereira, Claudia Mesquita, Silverio
  Visacro}, {Assessing currents of upward lightning measured in tropical
  regions}, Atmos. Res. 149 (2014) 324 -- 332.

\bibitem{Uman_and_Orville}
{Martin A. Uman and Richard E. Orville}, {The Opacity of Lightning}, J.
  Geophys. Res. 70 (1965) 5491--5497.

\end{thebibliography}

\end{document}